\documentclass[letterpaper]{article}

\usepackage[T1]{fontenc}

\usepackage{geometry}
\usepackage{setspace}
\usepackage[colorlinks=true, linkcolor=blue, citecolor=blue]{hyperref}
\usepackage[superscript]{cite}
\usepackage{hyperref}
\usepackage{graphicx}
\usepackage{float}
\newfloat{scheme}{htbp}{los}
\floatname{scheme}{Scheme}
\floatname{chart}{Chart}
\newfloat{graph}{htbp}{loh}
\usepackage{comment}
\usepackage{chemformula} % Formulas using \ch{}
\usepackage[version = 4]{mhchem} % Formulas using \ce{}
\usepackage{bm}
\usepackage{authblk}
\author[1,2]{Yu Tian}
\author[1]{Nuo Wang}
\author[1,2]{Qi Liu}
\author[3]{Shuyuan Xiao}
\author[3]{Tingting Liu}
\author[4]{Olivier J. F. Martin}
\author[1,2,5,6,7*]{Ying Gu}

\affil[1]{State Key Laboratory of Artificial Microstructure and Mesoscopic Physics $\&$ Department of Physics, Peking University, Beijing 100871, China}
\affil[2]{Frontiers Science Center for Nano-optoelectronics $\&$  Collaborative Innovation Center of Quantum Matter $\&$ Beijing Academy of Quantum Information Sciences, Peking University, Beijing 100871, China}
\affil[3]{School of Information Engineering, Nanchang University, Nanchang 330031, China}
\affil[4]{Nanophotonics and Metrology Laboratory, Swiss Federal Institute of Technology Lausanne (EPFL), Lausanne CH-1015, Switzerland}
\affil[5]{Collaborative Innovation Center of Extreme Optics, Shanxi University, Taiyuan, Shanxi 030006, China}
\affil[6]{Peking University Yangtze Delta Institute of Optoelectronics, Nantong 226010, China}
\affil[7]{Hefei National Laboratory, Hefei 230088, China}

\title{Wavefront Control and Intensity Modulation of Third Harmonic Generation in Nonlocal Metasurfaces}
\date{*Email: ygu@pku.edu.cn}

\begin{document}

\maketitle

\begin{abstract}

Metasurfaces have emerged as a promising platform for integrated nonlinear optics. Nonlocal metasurfaces enable high nonlinear conversion efficiency, while the local ones can offer versatile wavefront control, yet achieving both within a single metasurface remains challenging.
 	Here, using a nonlocal phase gradient metasurface, we firstly demonstrate efficient third harmonic generation (THG) with polarization-dependent wavefront control.
 	Leveraging nonlocal nonlinear geometric phase existing at resonance, the third harmonic light with distinct polarizations is deflected into $\pm$\,2nd  and $\pm $\,4th diffraction orders, simultaneously achieving a conversion efficiency up to $1.45\times 10^{-4}$ under a pump intensity of $1\,\mathrm{GW/cm^2}$.
 	Then, by introducing a secondary fundamental beam, whose generated third harmonic light overlaps with that of the first beam, the intensity modulation of THG is obtained.
 	The THG efficiency can be tuned from $3.9\times 10^{-9}$ to $5.5\times 10^{-3}$ by varying the relative phase, polarization and intensity of two fundamental beams. Through utilizing the advantages of both local and nonlocal metasurfaces, our results effectively pave the way to on-chip nonlinear photonic devices and signal processing.

\end{abstract}

\section*{Keywords}

\textit{nonlocal metasurfaces, quasi bound states in the continuum, third harmonic generation, high conversion efficiency, wavefront control, intensity modulation}

\section{Introduction}

Metasurfaces, a class of two-dimensional artificial materials, offer exceptional control over amplitude, phase, polarization and other properties of light at the micro/nano scale \cite{Sci2011 kaishan,Rep2016 zongshu, liuqi OE}, enabling various classical and quantum applications\cite{Sci 2018, liuqi APN, liuqi Light}.
By harnessing near-field enhancement, resonant metasurfaces further facilitate frequency-domain manipulation for nonlinear frequency conversion, which makes them promising candidates for integrated nonlinear optics \cite{Nature2014 jinshuzengqiang,Natrev2017 zongshu,ACSP2021 zongshu}.
To date, a variety of nonlinear processes, including harmonic generation and frequency mixing, have been demonstrated in metasurfaces with plasmonic resonances or Mie modes  \cite{Nano2016 jiezhizengqiang,NC2018 jiezhizengqiang,NN2015 jinshuzengqiang}.
For instance, third harmonic generation (THG) has been reported on silicon metasurfaces with conversion efficiencies exceeding $10^{-6}\sim10^{-5}$, nearly a four-order-of-magnitude enhancement over unpatterned silicon films \cite{Nano2014 Mie,Nano2015 Fano,ACAN2017 Mie}.
Despite these advances, the overall nonlinear conversion efficiencies of these metasurfaces remain relatively low, primarily because they operate under local modes with limited quality factors $(Q<10^{2})$.
To enhance the conversion efficiency, nonlocal metasurfaces, manifested as strong near-field enhancement and higher quality factors $(Q>10^{3})$ \cite{APL2023 zongshu,PR2023 zongshu}, have been introduced into nonlinear optics.
Among them, metsurfaces supporting quasi-bound states in the continuum (q-BICs) have emerged as an important branch, which can principally provide nearly infinite quality factors through tailoring structural symmetry and geometric perturbations \cite{Natrev2016 BICzongshu,PRL2018 BIC}, thereby boosting nonlinear processes \cite{PR2023 zongshu}.
Leveraging q-BICs, silicon metasurfaces have realized highly efficient THG with conversion efficiencies up to $10^{-4}$ \cite{PRL2019 BIC,OE2022 xiaoshuyuan,Nano xiaoshuyuan}, corresponding to nearly a two-order-of-magnitude improvement relative to their local-mode counterparts.
Beyond THG, nonlocal metasurfaces have also demonstrated higher-order nonlinear processes \cite{PRR2019 HHG,ACSP2022 HHG}, as well as quantum nonlinear phenomena like spontaneous parametric down-conversion \cite{AP2021 SPDC,Sci2022 SPDC}, all of which disclose their emerging prospects for advanced nonlinear photonics.

Despite the remarkable nonlinear enhancement provided by nonlocal metasurfaces, another key requirement is to simultaneously tailor the wavefront of generated harmonic light.
While in local metasurfaces, harmonic wavefront can be readily tailored through independently engineering nonlinear responses of individual unit cells with high precision, facilitating nonlinear diffraction \cite{Nano2019 yanshe,Nano2024 yanshe}, nonlinear holography \cite{Nano2018 quanxi,Nano2019 quanxi}, vortex beam generation \cite{Nano2018 woxuan}, and others.
However, achieving nonlinear wavefront control in nonlocal metasurfaces remains challenging, because the response of each unit cell is strongly influenced by that of others, which makes it difficult to engineer the nonlinear response of individual unit cells independently. 
Recently, nonlocal metasurfaces encoded with geometric phase gradient, termed as nonlocal phase gradient metasurfaces (NPGMs), have been proposed for spectrally selective wavefront manipulation in linear optics \cite{PRB2020 NPGM,PRL2020 NPGM,AP2021 NPGM}.
By extending such geometric phase from linear to nonlinear regime, it becomes possible to simultaneously achieve high conversion efficiency and flexible wavefront control on a single NPGM --- one of the primary objectives of our research.
Furthermore, if established, the capability for nonlinear wavefront control also supports interference among fundamental beams as well as among harmonic light, which offers opportunities for all-optical modulation of nonlinear responses --- another objective of our study.

The modulation of harmonic light in nonlinear metasurfaces is vital for signal processing and optical switch \cite{AP2019 zongshu,ACSP2024 zongshu}.
Typically, all-optical modulation is attained by introducing an additional control beam to tune the refractive index of nonlinear materials, which has been implemented for regulating the conversion efficiencies of second harmonic and third harmonic (TH) \cite{ACSP2016 kaiguan,ACSN2021 kaiguan,Nano2024 kaiguan}.
Nevertheless, these approaches predominantly rely on third-order nonlinear processes, like Kerr effect and two-photon absorption, whose nonlinear coefficients are remarkably low in many materials, thereby limiting the universality and depth of current modulation methods.
Alternatively, all-optical modulation can also be obtained through interference, which tailors the electric field distribution within the nonlinear materials.
This strategy has been implemented in conventional optical structures \cite{PRL2018 chuantong,PRR2025 chauntong}, facilitating modulation without dependence on a specific nonlinear coefficient.
However, leveraging interference to regulate conversion efficiencies of nonlinear metasurfaces remains largely unexplored.
By exploiting wavefront control capability of the NPGM, we reveal that the THG efficiency can be efficiently modulated by changing incident conditions of the secondary fundamental beam.

\begin{figure}[b!]
	\centering
	\setlength{\abovecaptionskip}{0pt}
	\includegraphics[width=1.0\textwidth]{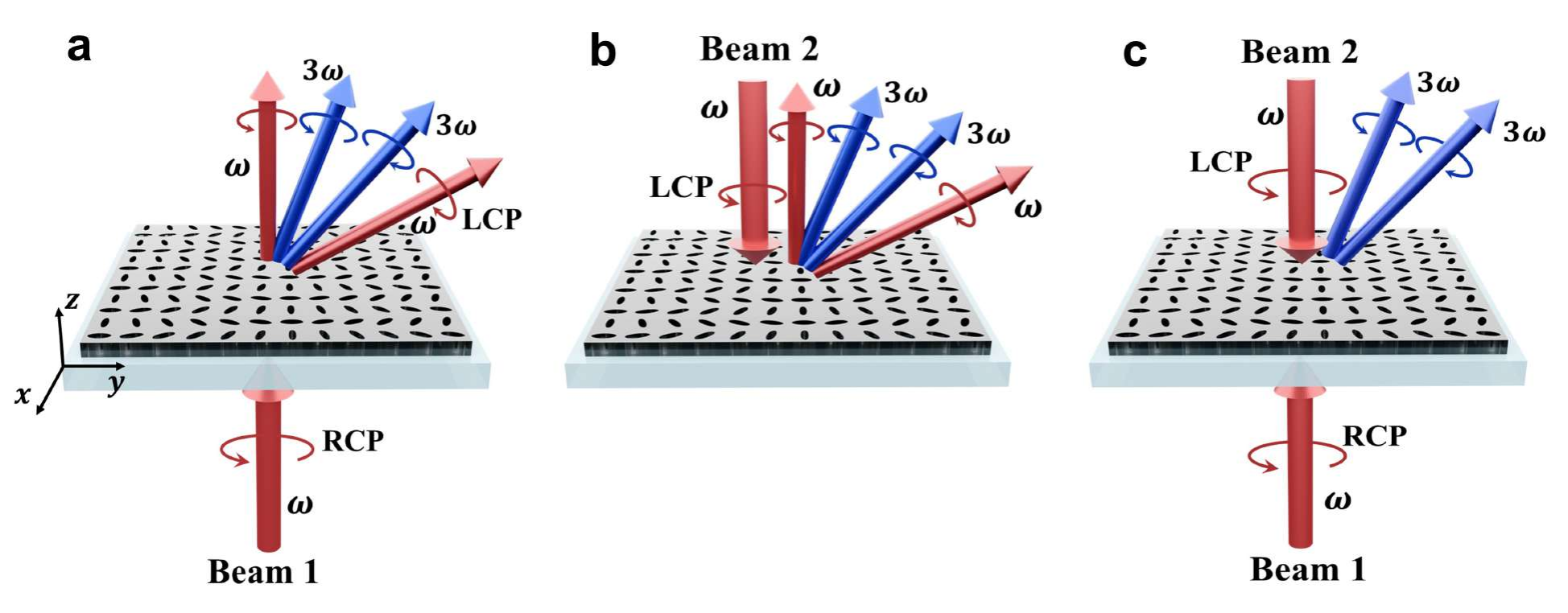}
	\caption{Schematic of wavefront control and intensity modulation of THG on a single NPGM. (a, b) A fundamental beam of (a) RCP light incident from substrate side or (b) LCP light incident from air side is incident into the NPGM. The generated TH light in the air side is deflected into $+$\,2nd and $+$\,4th diffraction orders. %For the fundamental beam of LCP light incident from substrate side or RCP light incident from air side, their TH light outputs from $-$\,2nd and $-$\,4th diffraction orders. 
	(c) Intensity modulation of THG on the same NPGM when Beam 1 and Beam 2 are incident simultaneously. Their generated TH light overlaps, leading to the modulation of THG efficiency with a near-unity depth.}
	\label{fig1}
\end{figure}

Here, we firstly demonstrate efficient THG with wavefront control on a single NPGM. The silicon-based NPGM supports q-BIC with a quality factor of approximately 120 and strong near-field enhancement. Building on this, a THG efficiency up to $1.45\times 10^{-4}$ is realized at a pump intensity of $1\,\mathrm{GW/cm^2}$. More importantly, a nonlocal nonlinear geometric phase can be imparted to THG at q-BIC through tuning the rotation angles of unit cells. This phase amounts to $4 \theta \sigma$ under co-polarized condition and $8 \theta \sigma$ under cross-polarized condition, which is exactly twice that in local metasurfaces\cite{Natrev2017 zongshu}, where $\theta$ denotes the rotation angles of unit cells and $\sigma=+1/-1$ represents the right/left-circularly-polarized (RCP/LCP) fundamental light, respectively. Exploiting our proposed nonlocal nonlinear geometric phase, polarization-dependent wavefront control is implemented. Under RCP fundamental beam incident, the generated TH light is deflected into the $+$\,2nd and $+$\,4th diffraction orders (\autoref{fig1}a), while it will be emitted into $-$\,2nd and $-$\,4th diffraction orders when the incident light is LCP. Then, by introducing a secondary fundamental beam, whose generated TH light overlaps with that of the primary beam (\autoref{fig1}b), the interference between two fundamental beams will strongly modulate the THG efficiency (\autoref{fig1}c).	With varying relative phase, polarization and intensity, the conversion efficiency can be dynamically tuned from $3.9\times 10^{-9}$ to $5.5\times 10^{-3}$, corresponding to a near-unity modulation depth. Our results overcome key limitations of the simultaneous realization of high conversion efficiency and flexible wavefront control, which pave the way toward compact and multifunctional nonlinear optical devices.	Furthermore, the interference-based modulation of nonlinear conversion efficiency holds promising prospects for on-chip nonlinear signal processing and optical switch.

\section*{Results and Discussion}

\subsection{1. Wavefront control of THG in NPGM}

\begin{figure}[b!]
	\centering
	\includegraphics[width=0.75\textwidth]{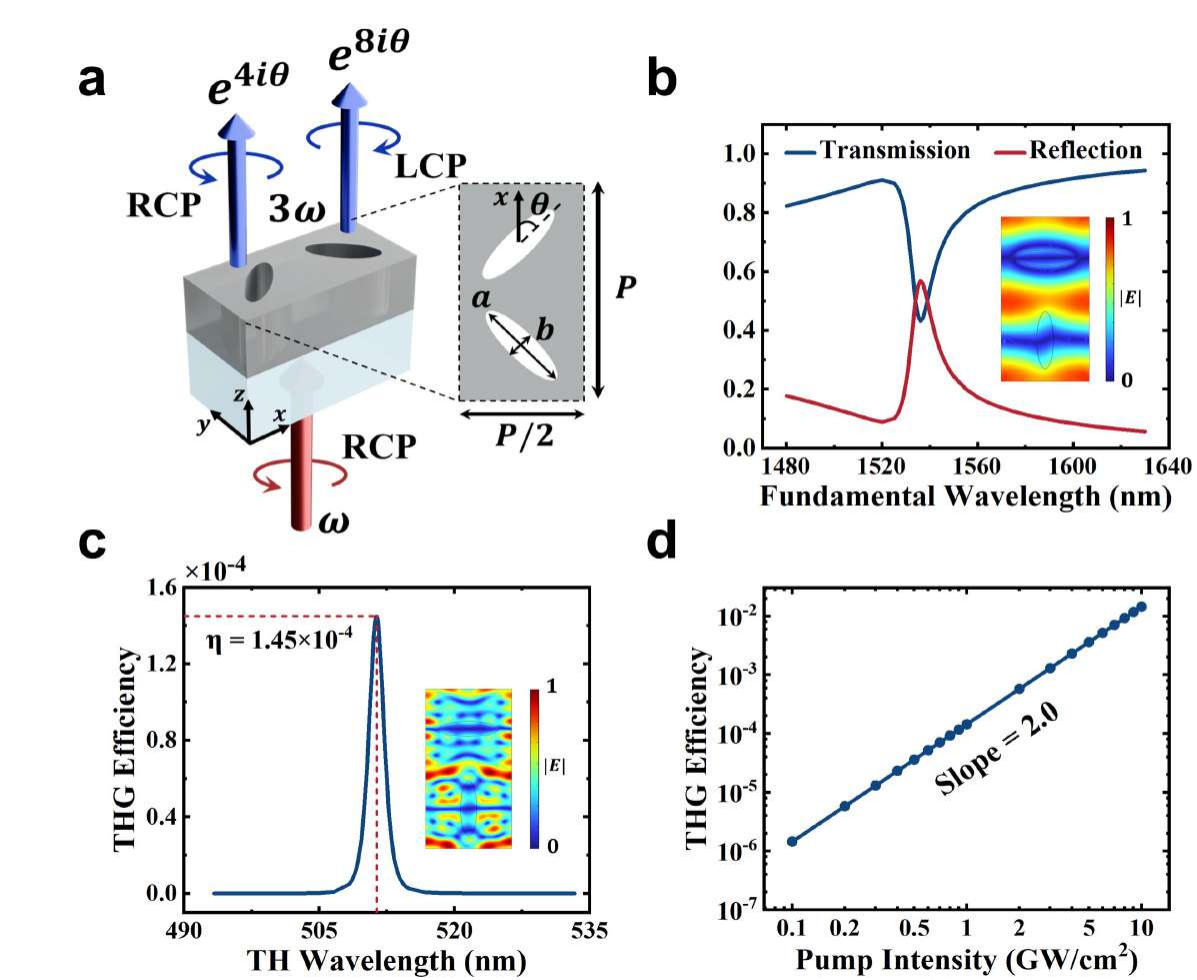}
	\caption{Enhanced THG within the unit cell of NPGM (a) The schematic of THG process in the periodic unit cell.  (b) The transmission/reflection spectra of unit cell with its electric field distribution in the xy-plane at q-BIC. (c) The THG efficiency of unit cell as a function of TH wavelength under a pump intensity of $1\,\mathrm{GW/cm^2}$, with the electric field distribution in the xy-plane at the wavelength of 511.3 nm. (d) The relation of THG efficiency with the pump intensity. Here, a fundamental beam of RCP light is incident into the unit cell normally from substrate.}
	\label{fig2}
\end{figure}

Now, we start to investigate efficient THG with wavefront control on the NPGM.
Such NPGMs have been reported to tailor the wavefront of linear optical fields, accompanied by strong near-field enhancement\cite{PRB2020 NPGM,PRL2020 NPGM,AP2021 NPGM}.
By applying it to the nonlinear optics, it is expected that high conversion efficiency and flexible wavefront manipulation of nonlinear optical fields can be attained on a single NPGM.
In this section, we firstly examine the enhanced THG process, whose efficiency is significantly improved compared with that of local metasurfaces\cite{Nano2014 Mie,ACAN2017 Mie}.
Notably, NPGM also provides the nonlocal nonlinear geometric phase for polarization-dependent wavefront control of TH light only at q-BIC.
Therefore, high nonlinear conversion efficiency and versatile wavefront engineering can be simultaneously obtained on a single NPGM, which provides promising prospects for on-chip applications of nonlinear metasurfaces.

{\normalsize \textbf{1.1 Linear and nonlinear responses of unit cell.}}
Consider the periodically arranged unit cell consisted of a silicon slab (gray part) etched with elliptical holes placed on a glass substrate (light blue part), as illustrated in \autoref{fig2}a. The refractive index of silicon is taken from experimental data (See Supporting Information Section S1)\cite{book}, while that of glass substrate is fixed as 1.5. The unit cell has a thickness of 260 nm and a period constant of P = 750 nm. Besides, the rotation angle of elliptical holes is fixed as $\theta =0^\circ$ with major axis of $a=$ 260 nm and minor axis of $b=$70 nm. For THG process in silicon unit cells, the nonlinear polarization can be expressed as $\bm{P}^{3\omega}=\varepsilon_{0} \chi_{3} (\bm{E}^{\omega} \cdot \bm{E}^{\omega}) \bm{E}^{\omega}$, where $\varepsilon_{0}$ is the vacuum permittivity, $\chi_{3}=2.45\times 10^{-19}\,\mathrm{ m^{2}/V^{2}}$ is the third-order susceptibility of silicon\cite{chi}, and ${\bm{E}^{\omega}}$ is the electric field at the fundamental wavelength.
Here, the TH conversion efficiency is defined as $\eta =P_{TH}/P_{F}$, where $P_{TH}$ denotes the TH power radiated from the air side, and $P_{F}$ refers to the pump power at the fundamental wavelength. The numerical analysis is performed using the finite element method solver in commercial software COMSOL Multiphysics in the frequency domain (further details are included in the Methods section). 

We begin by investigating linear and  nonlinear responses of the above unit cell. 
%Such periodic unit cells have been constructed to support q-BICs in the near-infrared wavelength range to spatially shape the output wavefront\cite{PRB2020 NPGM,PRL2020 NPGM,AP2021 NPGM}. 
It supports the q-BIC located at 1536 nm with a quality factor of approximately 120, as confirmed by the transmission and reflection spectra in \autoref{fig2}b. The q-BIC facilitates strong near-field enhancement within silicon materials (the inset of \autoref{fig2}b), which can be applied to enhance THG process. 
When the pump beam is RCP with an intensity of $1\,\mathrm{GW/cm^2}$, the conversion efficiency reaches its peak of $1.45\times 10^{-4}$ at the TH wavelength of 511.3 nm (\autoref{fig2}c), representing nearly a two-order-of-magnitude improvement compared to local metasurfaces\cite{Nano2014 Mie,Nano2015 Fano,ACAN2017 Mie}.
Consistent with the preceding discussion, THG predominantly occurs at the position of strong near-field enhancement at q-BIC (the inset of \autoref{fig2}c). 
Moreover, as the pump intensity increases, the THG efficiency increases linearly with a fitted slope of 2.0 with logarithmic coordinate (\autoref{fig2}d), confirming the characteristics of THG process\cite{Nano2024 yanshe}.
Additionally, TH light is also emitted from the substrate side, whose intensity is slightly lower than that of air side (See Supporting Information Section S2). Owing to their similar properties, we only focus on the intensity and wavefront of the air-side TH light. 
Therefore, efficient THG has been implemented at q-BIC with the consideration of unit cell in the NPGM.

{\normalsize \textbf{1.2 Nonlocal nonlinear geometric phase.}} 	In addition to THG enhancement, a nonlocal nonlinear geometric phase can be imparted to TH light. Specifically, when the fundamental beam at q-BIC is incident into  unit cells of the NPGM, the RCP and LCP components of generated TH light will carry geometric phases determined by the rotation angle $\theta$ of elliptical holes (\autoref{fig2}a). The relations between geometric phase and $\theta$ can be analytically obtained from the nonlinear polarization of THG in circularly polarized basis. Taking RCP light incidence as an example, the RCP and LCP components of nonlinear polarization can be expressed as
\begin{equation}
	P_R^{3\omega} = e^{i4\theta} \varepsilon_0 \chi_\text{RRRR} E_R^{\omega} E_R^{\omega} E_R^{\omega}, \quad P_L^{3\omega} = e^{i8\theta} \varepsilon_0 \chi_\text{LRRR} E_R^{\omega} E_R^{\omega} E_R^{\omega},
\end{equation}
 where $\chi_\text{LRRR}$ refers to the nonlinear coefficient that enables three RCP photons at frequency $\omega$ to generate one LCP photon at frequency $3\omega$.
As a result, when a beam of RCP light at q-BIC is incident into the unit cell with a rotation angle of $\theta$, the LCP component of TH light carries a phase of $8\theta$, while RCP component carries a phase of $4\theta$ (\autoref{fig2}a).
%, as confirmed by our numerical results (\autoref{fig3}c).
Conversely, when the incident light is LCP, the LCP component of TH light carries a phase of $-4\theta$, and the RCP component carries a phase of $-8\theta$ (see Supporting Information Section S3 for detailed derivation).

\begin{figure}[t!]
	\centering
	\setlength{\abovecaptionskip}{0pt}
	\includegraphics[width=1.0\textwidth]{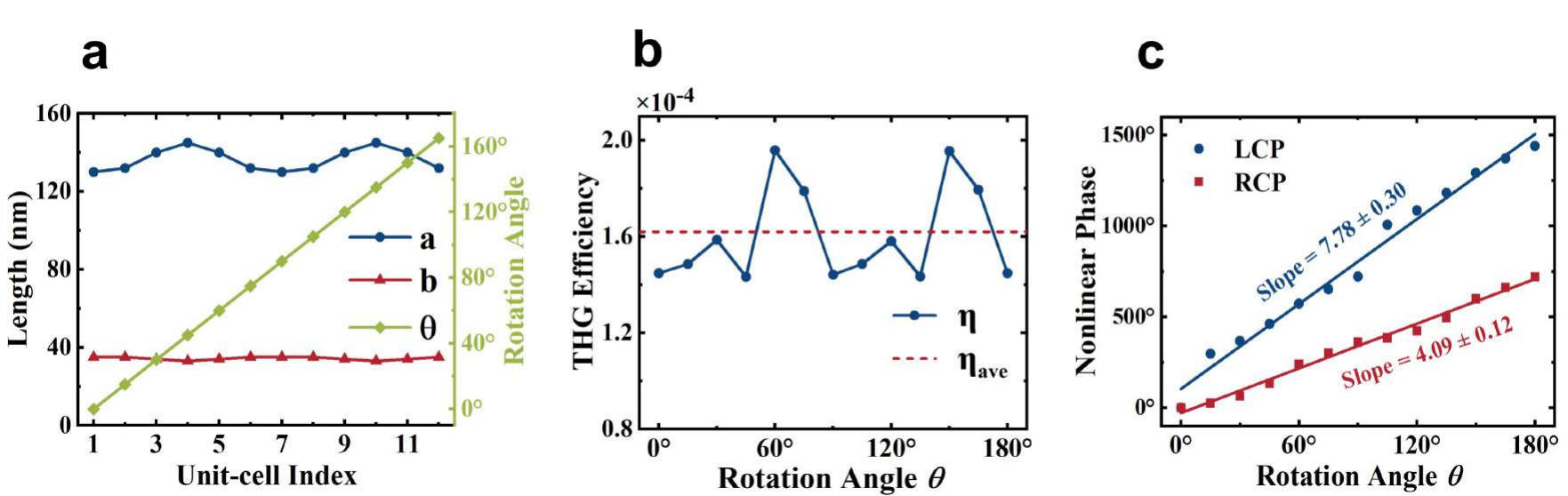}
	\caption{(a) The geometric parameters of unit cells in the NPGM. The major and minor axes of elliptical holes are adjusted with rotation angles to maintain the same q-BIC wavelength and similar THG efficiencies. (b) The THG efficiency and (c) nonlinear geometric phase of periodic unit cells with varying rotation angles. Here, the fundamental beam is RCP light with an intensity of $1\,\mathrm{GW/cm^2}$ to generate TH light of 511.3 nm.}
	\label{fig3}
\end{figure}

The above analysis about nonlocal nonlinear geometric phase can be verified through simulations. We first construct 12 types of unit cells with a relative rotation angle of $15^\circ$. These unit cells are the same in period constant and thickness, but the major axis $a$ and minor axis $b$ of their elliptical holes are adjusted with rotation angles accordingly (\autoref{fig3}a). This is to ensure all unit cells support the same wavelength of q-BIC as well as comparable THG efficiencies, as illustrated in \autoref{fig3}b. Only under these conditions, the far-field distribution of TH light is solely determined by the nonlocal nonlinear geometric phase encoded by unit cells, thereby allowing nonlinear wavefront manipulation. Then, the relations between noonlinear geometric phases of TH light and rotation angles $\theta$ are calculated. Under RCP light illumination at the q-BIC, when the rotation angles of unit cell changes by $\theta$, the nonlinear phase of RCP component of TH light varies by $4\theta$, and that of LCP component changes by $8\theta$ (\autoref{fig3}c), which are consistent with the above analytical results in Eq. (1). For LCP fundamental beam incidence, the conditions are similar. Therefore, the nonlocal nonlinear geometric phase is well established, which is the foundation of nonlinear wavefront engineering in the NPGM.

Notably, the proposed nonlocal nonlinear geometric phase is distinct from the previously reported nonlinear geometric phase in local metasurfaces in terms of response amplitude and operating bandwidth\cite{Natrev2017 zongshu,Nano2019 yanshe, Nano2019 quanxi, AOM 2020, Adv Sci 2022}. Typically, for linear/nonlinear processes in local metasurfaces, the geometric phases are $(n-1)\theta \sigma$ and $(n+1)\theta\sigma$ for co-polarization and cross-polarization conditions, respectively\cite{Natrev2017 zongshu}, where $n$ refers to the order of nonlinear processes, $\theta$ denotes the rotation angle of unit cells and $\sigma = +1/-1$ represents RCP/LCP fundamental beam. For linear process ($n=1$), they are $0$ and $2\theta\sigma$ for co-polarization and cross-polarization conditions, respectively\cite{liuqi OE}; for THG process ($n=3$), they are $2\theta \sigma$ for co-polarization condition and $4\theta \sigma$ for cross-polarization condition\cite{Nano2019 quanxi}. These are established by the fact that 
when the rotation angle of unit cell of local metasurfaces changes by $\theta$, the principal axis also rotates by $\theta$ accordingly. However, for the unit cell in nonlocal phase gradient metasurfaces, a rotation by $\theta$ leads to a $2\theta$ rotation of its principal axis\cite{PRB2020 NPGM}. As a result, the geometric phases for linear optical fields are $0$ and $4\theta\sigma$ under co and cross-polarization conditions, respectively\cite{PRL2020 NPGM,AP2021 NPGM}, which are twice those in local metasurfaces. Accordingly, the nonlocal nonlinear geometric phases for THG process are also twice those of local metasurfaces, which are $4\theta\sigma$ and $8\theta\sigma$ under co and cross-polarization conditions. These are also confirmed by the above analytical results in Eq. (1) and simulations. Moreover, in contrast to the broadband responses in local metasurfaces\cite{Natrev2017 zongshu,AOM 2020},  the nonlinear geometric phase in the NPGM only exists at q-BIC. Thus, the nonlocal nonlinear geometric phase enables the NPGM to engineer the wavefront of TH light at q-BIC.

\begin{figure}[h!]
	\centering
	\setlength{\abovecaptionskip}{7pt}
	\includegraphics[width=0.75\textwidth]{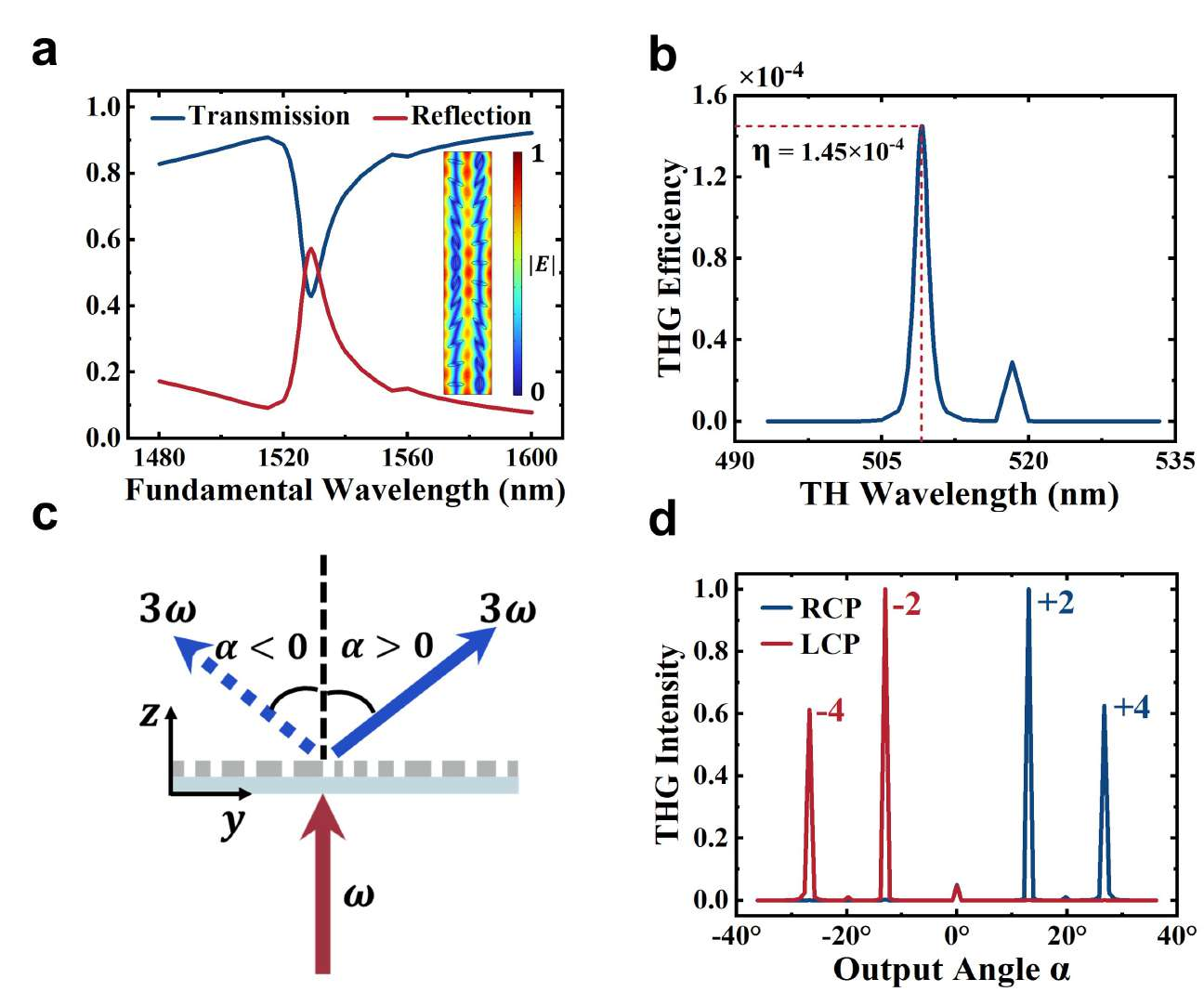}
	\caption{(a) The transmission/reflection spectra of NPGM with electric field distribution in the xy-plane at q-BIC. (b) The THG efficiency  of NPGM as a function of TH wavelength. (c) The schematic of output angle $\alpha$ of TH light. (d) The distribution of normalized THG intensity with RCP and LCP fundamental beams incidence from substrate side normally.}
	\label{fig4}
\end{figure}

{\normalsize \textbf{1.3 Wavefront control of THG.}}  To manipulate the wavefront of TH light, we firstly construct the NPGM composed of the above 12 types of unit cells (\autoref{fig1}a).
It supports the q-BIC located at 1529 nm with transmission and reflection spectra displayed in \autoref{fig4}a, which exhibits a slight blueshift relative to the condition of single-unit-cell periodic arrangement\cite{PRB2020 NPGM,PRL2020 NPGM,AP2021 NPGM}.
There is strong near-field enhancement within silicon materials at q-BIC (the inset of \autoref{fig4}a), thereby boosting the THG process.
As a result, when the incident fundamental beam is RCP with an intensity of $1\,\mathrm{GW/cm^2}$, the THG efficiency exceeds $\eta =1.45\times 10^{-4}$ at the TH wavelength of 509 nm, as illustrated in \autoref{fig4}b. Compared with local metasurfaces with phase gradients \cite{Nano2018 quanxi,Nano2019 quanxi,Nano2018 woxuan}, this is an improvement of more than two orders of magnitude. Meanwhile, the intensity of TH light emitting from substrate side is relatively lower than that of air side (See Supporting Information Section S2).
Therefore, by exploiting the NPGM, highly efficient THG process can be implemented at the q-BIC.

While enhancing THG efficiency, polarization-dependent wavefront control of TH light can be achieved at q-BIC. 
The THG intensity distribution among output angle $\alpha$ can be obtained through Fourier transform (See details in the Methods Section), where $\alpha$ is defined as the angle between the emission direction of TH light and +z axis (\autoref{fig4}c).
As illustrated in \autoref{fig4}d, with RCP fundamental beam incidence, the output angles of TH light are mainly $\alpha=13.1^\circ$ and $26.9^\circ$, corresponding to the $+$\,2nd and $+$\,4th diffraction orders at the wavelength of 509 nm, respectively. This nonlinear wavefront control originates from the nonlocal nonlinear geometric phases discussed above. According to Eq. (1) and numerical results in \autoref{fig3}c, the phases of TH light for co and cross-polarization conditions in the NPGM are $+4\theta$ and $+8\theta$, respectively, indicating that the TH light will output from $+$\,2nd and $+$\,4th diffraction orders, which is in good agreement with the numerical results in \autoref{fig4}d.
%提法有点问题
Meanwhile, the fundamental beam mainly outputs from directions with output angles of $0^\circ$ and $42.7^\circ$, corresponding to the 0th and $+$\,2nd diffraction orders at the wavelength of 1527 nm (see Supporting Information Section S4), which are also consistent with the reported geometric phases of $0$ and $4\theta$ for linear optical fields\cite{PRB2020 NPGM,PRL2020 NPGM,AP2021 NPGM}.
In contrast, when the incident fundamental light is LCP, the TH light is primarily emitted into $-$\,2nd and $-$\,4th diffraction orders (\autoref{fig4}d).
For fundamental beam with other polarizations, since it can be regarded as the superposition of RCP and LCP light, its TH light will output from the $\pm$\,2nd and $\pm$\,4th diffraction orders, simultaneously.
Thus, in the NPGM, the wavefront of TH light can be tailored at q-BIC with polarization-dependent and spectrally selective characteristics.

In a word, we have demonstrated highly efficient THG with polarization-dependent wavefront manipulation at q-BIC on a single NPGM, overcoming key limitations of the simultaneous realization of high conversion efficiency and wavefront control in nonlinear metaurfaces.
Building on these, more flexible wavefront control can be realized through NPGMs by designing nonlocal nonlinear geometric phase gradients. So it is expected that NPGM can facilitate more advanced functionalities, such as nonlinear holography \cite{Nano2018 quanxi,Nano2019 quanxi} and vortex beam generation \cite{Nano2018 woxuan}, which are essential for versatile nonlinear light sources.
In contrast to local metasurfaces with broadband responses \cite{Natrev2017 zongshu,AOM 2020}, NPGMs possess inherently narrowband phase responses. This allows spectrally selective wavefront engineering of nonlinear optical fields to promote the development of multifunctional nonlinear photonic devices.
Beyond classical nonlinear optics, the NPGM can also be  extended to facilitate quantum nonlinear processes accompanied by wavefront control, including spontaneous parametric down-conversion and four-wave mixing, which offers opportunities for the generation and manipulation of quantum states with larger entanglement and efficiency. The wavefront control capability of the NPGM also provide possibilities for intensity modulation of THG via interference, which we will discuss in the following section.

\subsection{2. Intensity Modulation of THG}

Here, we intend to investigate the intensity-modulation of THG via interference in the same NPGM. The discussions in previous section are focused on passive nonlinear responses with fixed functionalities once fabricated. All-optical modulation is crucial for on-demand applications of the NPGM, which can be realized through the interference between two fundamental beams. In this section, based on the polarization-dependent wavefront control for TH light, a secondary fundamental beam is introduced, whose TH light overlaps with that of the first fundamental beam (\autoref{fig1}b). By inputting these two fundamental beams simultaneously (\autoref{fig1}c), the THG efficiency can be tuned from $3.9\times 10^{-9}$ to $5.5\times 10^{-3}$ by adjusting the relative phase, intensity and polarization. The interference-based modulation of THG exhibits a near-unity modulation depth, which can be applied in all-optical switch and nonlinear signal processing.

{\normalsize \textbf{2.1 The mechanism for intensity modulation.}} 	We first elucidate the physical mechanism governing the modulation of THG efficiency through interference between two fundamental beams.
When only the first fundamental beam is incident, the induced electric field distribution in the NPGM is denoted as $\bm{E}_{1}^{\omega}=(E_{1x},E_{1y},E_{1z} )^{T}$, whose specific form is determined by its phase, intensity and polarization.
When only the second fundamental beam illuminates the NPGM, the induced electric field distribution is referred as $\bm{E}_{2}^{\omega}=(E_{2x},E_{2y},E_{2z} )^{T}$.
If these two fundamental beams are incident simultaneously, the total electric field distribution in the NPGM becomes the coherent superposition of $\bm{E}_{1}^{\omega}$ and $\bm{E}_{2}^{\omega}$, expressed as $\bm{E}_{s}^{\omega}=\bm{E}_{1}^{\omega}+\bm{E}_{2}^{\omega}=(E_{1x}+E_{2x},E_{1y}+E_{2y},E_{1z}+E_{2z})^{T}$.
Accordingly, the nonlinear polarization for THG in the NPGM is given by
\begin{equation}
	\bm{P}_{s}^{3\omega}=\varepsilon_{0} \chi_{3} (\bm{E}_{s}^{\omega} \cdot \bm{E}_{s}^{\omega}) \bm{E}_{s}^{\omega},
\end{equation} 
where $\varepsilon_{0}$ is the vacuum permittivity, $\chi_{3}=2.45\times 10^{-19}\,\mathrm{ m^{2}/V^{2}}$ is the third-order susceptibility of silicon\cite{chi}.
So, the THG efficiency is determined by the total electric field distribution $\bm{E}_{s}^{\omega}$ formed through the interference between two fundamental beams.
By varying the incident conditions of secondary fundamental beam, $\bm{E}_{s}^{\omega}$ can be substantially tailored, thereby facilitating efficient modulation of THG efficiency.
Therefore, the effective interference between two fundamental beams can support the intensity modulation of THG.
\begin{comment}
\begin{figure}[h!]
	\centering
	\setlength{\abovecaptionskip}{0pt}
	\includegraphics[width=0.9\textwidth]{Fig5(1).eps}
	\caption{Schematic of the mechanism for THG intensity modulation in the NPGM. (a) Beam 1: RCP fundamental light with an electric amplitude of $E_1$ and phase of $\varphi _1$ incident from substrate side. Its induced electric field distribution at fundamental wavelength in the NPGM is $\bm{E}_1^{\omega}$, generating TH light of $\bm{E}^{3\omega}$ emitted into $+$\,2nd and $+$\,4th diffraction orders. (b) Beam 2: fundamental light ($E_2$, $\varphi_2$) with arbitrary polarization incident from air side. Its induced electric field distribution at fundamental wavelength in the NPGM is $\bm{E}_2^{\omega}$, generating TH light $\bm{E}^{3\omega}$ emitted into $\pm$2nd and $\pm$4th orders. (c) Intensity modulation: simultaneous incidence of Beam 1 and 2 leads to total induced fundamental field $\bm{E}_s^{\omega} = \bm{E}_1^{\omega} + \bm{E}_2^{\omega}$; varying relative phase, intensity, and polarization adjusts $\bm{E}_s^{\omega}$, modulating THG intensity. }
\end{figure}
\end{comment}

In the NPGM, the effective interference occurs with overlapping of TH light generated by two fundamental beams, as schematically illustrated in \autoref{fig1}.
As discussed in the previous section, for the RCP/LCP fundamental beam incidence from substrate side, the TH light is predominantly emitted into $+$\,2nd/$-$\,2nd and $+$\,4th/$-$\,4th diffraction orders.
Conversely, when RCP/LCP fundamental light is incident from air side, the TH light primarily outputs from $-$\,2nd/$+$\,2nd and $-$\,4th/$+$\,4th diffraction orders.
Considering these, we fix the first fundamental beam (Beam 1) as RCP light incidence from substrate side with an intensity of $1\,\mathrm{GW/cm^2}$, while the second fundamental beam (Beam 2) is incident from air side with varying phase, intensity and polarization.
Under this configuration, when Beam 2 is LCP, strong interference occurs, since its generated TH light overlaps with that produced by Beam 1 (\autoref{fig1}c). When Beam 2 is RCP, the interference will almost disappear.
Consequently, by varying the phase, intensity and polarization  of Beam 2, the intensity modulation of THG can be obtained in the same NPGM.

\begin{figure}[t!]
	\centering
	\setlength{\abovecaptionskip}{5pt}
	\includegraphics[width=0.75\textwidth]{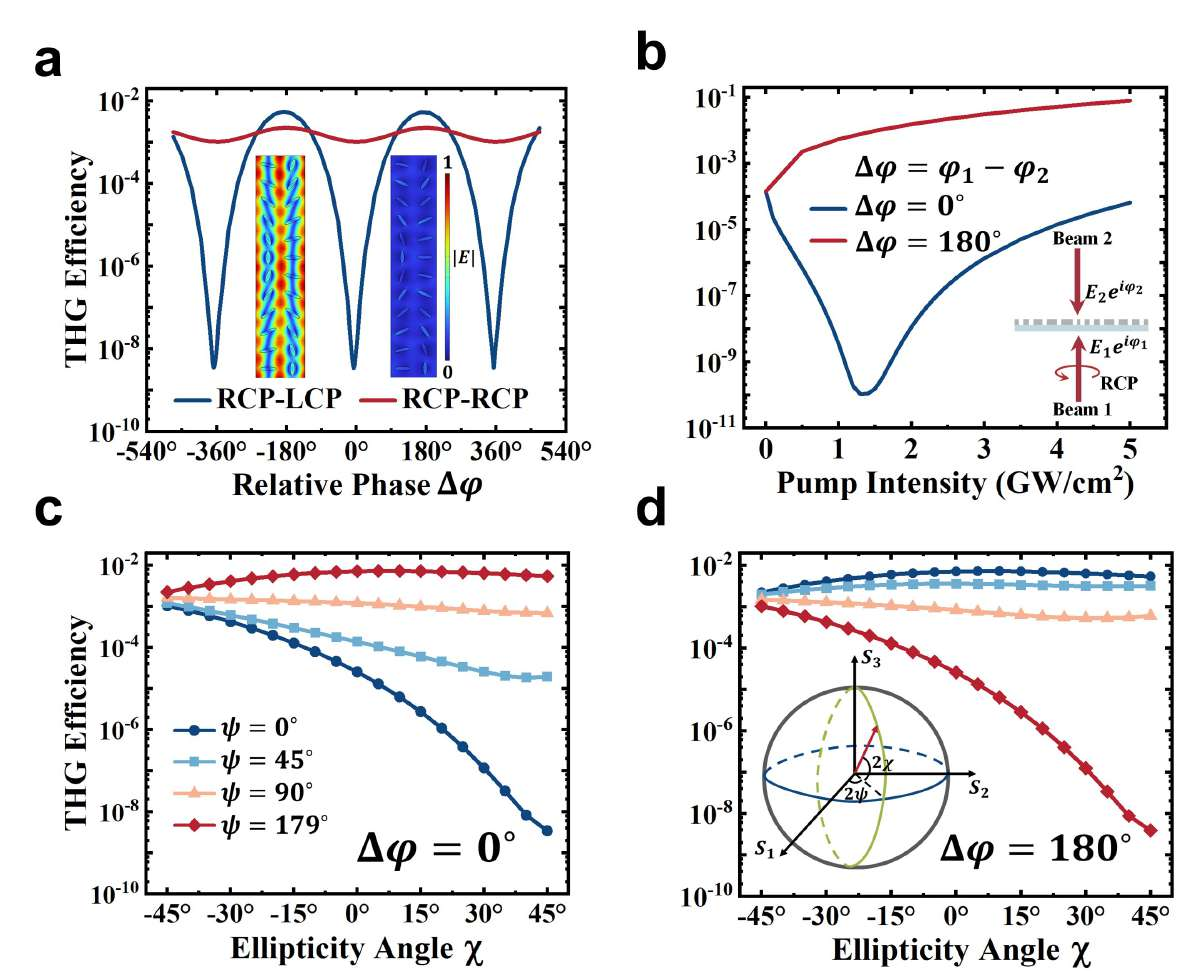}
	\caption{The intensity modulation of THG efficiency through varying (a) relative phase $\Delta \varphi$, (b) pump intensity of Beam 2, and polarization of Beam 2 with (c) $\Delta \varphi=0^\circ$ and (d) $\Delta \varphi=180^\circ$. The insets in (a) are the electric field distribution of fundamental beam in the xy-plane at $\Delta \varphi=180^\circ$ (left) and $\Delta \varphi=0^\circ$ (right), which exhibits large near-field enhancement under constructive interference, while approaches zero under destructive interference. The inset in (d) is the physical meanings of orientation angle $\psi$ and ellipticity angle $\chi$ on the Poincaré sphere.}
	\label{fig5}
\end{figure}

{\normalsize \textbf{2.2 The demonstration of intensity modulation.}} 	By adjusting the phase of Beam 2, the THG efficiency can be manipulated with a near-unity modulation depth. Here, Beam 2 is RCP or LCP light with an intensity of $1\,\mathrm{GW/cm^2}$ and a relative phase $\Delta \varphi$ with respect to Beam 1, where $\Delta \varphi$ is defined as the difference in propagation phases between Beam 1 and Beam 2, namely, $\Delta \varphi=\varphi_{1}-\varphi_{2}$ (the inset of \autoref{fig5}b). As shown in \autoref{fig5}a, when Beam 2 is LCP, the THG efficiency can be tuned from $3.9\times 10^{-9}$ to $5.5\times 10^{-3}$, spanning over six orders of magnitude. This represents a near-unity modulation depth, which is superior to existing results relying on Kerr effect and two-photon absorption\cite{ACSP2016 kaiguan,ACSN2021 kaiguan,Nano2024 kaiguan}.
Notably, the THG efficiency is minimum at $\Delta \varphi=0^\circ$, while it is maximum at $\Delta \varphi=180^\circ$. This is due to the coupling coefficient between q-BIC and Beam 1 is opposite to that of Beam 2 (see Supporting Information Section S5). When Beam 2 is RCP, the relation between THG efficiency and $\Delta \varphi$ is completely different. As $\Delta \varphi$ is varied, the THG efficiency remains nearly constant, as shown by the red curve in \autoref{fig5}a. This is because the generated TH light from the two fundamental beams has nearly no overlap, leading to the absence of interference.
Therefore, the intensity modulation of THG in the NPGM can be achieved by changing the phase of Beam 2 of LCP.

Then, based on this, we explore the modulation of THG efficiency by varying the intensity of Beam 2 with LCP state (\autoref{fig5}b). As the intensity of Beam 2 increases at $\Delta \varphi=0^\circ$, the THG efficiency initially decreases and then increases, with a minimum efficiency of $\eta = 1.1\times 10^{-10}$ under an intensity of $1.4\,\mathrm{GW/cm^2}$.
This behavior results from destructive interference between two fundamental beams, which has $|\bm{E}_{s}^{\omega}|=\bigl|\,|\bm{E}_{1}^{\omega}|-|\bm{E}_{2}^{\omega}|\,\bigr|$. 
In contrast, as the intensity of Beam 2 increases at $\Delta \varphi=180^\circ$, the THG efficiency increases continuously, since the two beams are under constructive interference, namely, $|\bm{E}_{s}^{\omega}|=\bigl|\,|\bm{E}_{1}^{\omega}|+|\bm{E}_{2}^{\omega}|\,\bigr|$.
Thus, by properly adjusting the intensity of Beam 2, we can also engineer the THG efficiency in the NPGM.

Furthermore, we systematically analyze influences of the polarization of Beam 2 on THG efficiency. For Beam 2 propagating along the -z axis, its Jones vector can be written as
\begin{equation}
	\begin{bmatrix} E_x \\ E_y \end{bmatrix} = \begin{bmatrix} \cos \chi \cos \psi - i \sin \chi \sin \psi \\ \cos \chi \sin \psi + i \sin \chi \cos \psi \end{bmatrix},
\end{equation}
where $\psi\in[0^\circ,180^\circ)$ is the orientation angle, and $\chi\in[-45^\circ,45^\circ]$ is the ellipticity angle.
When $\chi=45^\circ$ and $-45^\circ$, it corresponds to LCP and RCP states with Jones vectors of
$|L_- \rangle = {1}/{\sqrt{2}} \begin{bmatrix} 1 , i \end{bmatrix}^T$
and 
$|R_- \rangle = {1}/{\sqrt{2}} \begin{bmatrix} 1 , -i \end{bmatrix}^T$, respectively.
When $\chi=0^\circ$, it represents linear polarization, while other values of $\chi$ correspond to elliptical polarization.
For Beam 2 with arbitrary polarization $\lvert a \rangle$, it can be decomposed into RCP and LCP components as
$\lvert a \rangle = m \lvert R_- \rangle + n \lvert L_- \rangle$,
where the coefficients are given by
$m=\cos(\chi+45^\circ)\mathrm e^{i\psi}$ and
$n=\sin(\chi+45^\circ)\mathrm e^{-i\psi}$,
satisfying $\lvert m\rvert^2+\lvert n\rvert^2=1$.
The amplitudes and phases of coefficients $m$ and $n$ can be adjusted by varying orientation angle $\psi$ and ellipticity angle $\chi$.
According to the above discussion, the LCP component of Beam~2 can effectively interfere with Beam~1, whereas the RCP component does not contribute to the interference. Therefore, the coefficient $n$ of LCP component of Beam 2 can be modulated by adjusting its polarization, thereby modulating the THG efficiency, as illustrated in \autoref{fig5}c,d.

When the initial relative phase is $\Delta \varphi = 0^\circ$, destructive interference occurs between Beam 1 and Beam 2, leading to the minimum THG efficiency.
As illustrated in \autoref{fig5}c, the THG efficiency reaches its minimum value of $\eta =3.9\times 10^{-9}$ with $\psi = 0^\circ, \chi = 45^\circ$, corresponding to Beam 2 as LCP light with an additional phase of $0^\circ$.
Conversely, the THG efficiency peaks at $\eta =5.3\times 10^{-3}$ when $\psi = 179^\circ,\chi = 45^\circ$, corresponding to Beam 2 as LCP light with an additional phase of $179^\circ$. If $\chi$ takes other values, the THG efficiency increases with orientation angle $\psi$. This is because under these conditions, the amplitude of LCP component coefficient $n$ remains constant, while its phase gradually increases with $\psi$, shifting from destructive interference to constructive interference.
On the other hand, when $\psi$ is fixed, an increase in $\chi$ leads to a larger amplitude of LCP component, which results in a decrease in THG efficiency for $\psi < 90^\circ$ (destructive interference), and an increase for $\psi > 90^\circ$ (constructive interference).
Thus, the THG efficiency, spanning over six orders of magnitude, exhibits remarkable sensitivity to the polarization of Beam 2, enabling polarization-controlled intensity switch with a near-unity modulation depth.
When the initial relative phase is $\Delta \varphi = 180^\circ$, constructive interference occurs between Beam 1 and Beam 2, leading to the maximum THG efficiency.
As illustrated in \autoref{fig5}d, the polarization-dependent modulation is exact opposite of that observed for $\Delta \varphi = 0^\circ$ in \autoref{fig5}c.
The THG efficiency peaks at $\eta =5.5\times 10^{-3}$ for $\psi = 0^\circ,  \chi = 45^\circ$, while it reaches its minimum value of $\eta =3.5\times 10^{-9}$ at $\psi = 179^\circ,\chi = 45^\circ$, also spanning over six orders of magnitude.
Therefore, we demonstrate near-unity modulation of THG efficiency through adjusting the polarization of Beam 2, which holds significant potential for applications in polarization-controlled switch in nonlinear optics.

\section*{Conclusions}

In conclusion, we have achieved an efficient THG process with polarization-dependent wavefront control on a single NPGM, and further demonstrated the intensity switch of THG.
By exploiting q-BIC, the THG efficiency of the silicon-based NPGM exceeds $10^{-4}$  under a pump intensity of $1\,\mathrm{GW/cm^2}$, representing an enhancement of nearly two orders of magnitude relative to local metausurfaces.
Simultaneously, nonlocal nonlinear geometric phase can be obtained at q-BIC for polarization-dependent wavefront control for TH light.
Furthermore, through introducing a secondary fundamental beam, intensity switch of THG is achieved via linear interference.
The THG efficiency can be dynamically tuned from $10^{-9}$ to $10^{-3}$ by adjusting the relative phase, intensity and polarization of the second beam, spanning over six orders of magnitude with a near-unity modulation depth.
These results address a fundamental challenge in nonlinear metasurfaces, which is to realize high conversion efficiency and flexible wavefront control simultaneously, thereby providing a viable route toward multifunctional nonlinear optical devices.
Moreover, the demonstrated interference-based, near-unity modulation of nonlinear conversion efficiency highlights the potential of this platform for nonlinear optical switch and on-chip signal processing.

\section*{Methods}
\subsection{Numerical simulation}
The linear and nonlinear responses of the NPGM is calculated through COMSOL Multiphysics. The general simulation model is an infinite periodic surface consisted with a silicon slab with elliptical holes placed on a glass substrate, as shown in \autoref{fig1} and \autoref{fig2}. The periodicity of structure is modeled by introducing periodic boundary conditions along $x$ and $y$ directions. In the top and bottom surface of the model, perfect match layer (PML) are applied for open boundary conditions. The periodic ports are utilized to serve as the input optical fields propagating along $+z$ or $-z$ directions. The refractive index of amorphous silicon is taken from experimental date considering the dispersion and dissipation\cite{book}, which is displayed in the Supporting Information Section S1, while those of substrate and air are fixed as 1.5 and 1.0, respectively. In simulations, the THG is taken into account through the nonlinear polarization $\mathbf{P}^{3\omega}=\varepsilon_{0} \chi_{3} (\mathbf{E}^{\omega} \cdot \mathbf{E}^{\omega}) \mathbf{E}^{\omega}$, where $\chi_{3}=2.45\times 10^{-19}\,\mathrm{ m^{2}/V^{2}}$ is obtained through experimental data\cite{chi}. During this process, the backward frequency conversion is neglected. Therefore, two frequency-domain solvers are required in the calculations, which correspond to the fundamental and TH wavelengths, respectively. The optical field distribution obtained by the first solver serve as the input of the second solver. For the second solver, continuous boundary conditions are performed along $x$ and $y$ directions, while PML condition is used along $z$ direction. Then, the surface integral of TH light power flux over the top and bottom surfaces of model can be conducted to measure the output power of TH light. To measure the output angles of TH light, Fourier transform is performed through COMSOL Multiphysics. An $x-y$ cross-section sufficiently far from the NPGM is obtained and then replicated along $x$ and $y$ directions to acquire the far field distribution of TH light in the spatial domain. Then, perform the Fourier transform on it and extract the data corresponding to $k_x=0$.

\section*{Supporting information}

\section{Notes}
The authors declare no competing financial interest.

\section*{Acknowledgements}

This work is supported by the National Natural Science Foundation of China under Grant No. U25D9003 and No. 12474370 and the Quantum Science and Technology-National Science and Technology Major Project No. 2021ZD0301500.

\end{document}